\documentclass[runningheads]{llncs}
\usepackage[T1]{fontenc}
\usepackage{graphicx}
\usepackage{amsfonts}
\usepackage{amsmath}
\usepackage{tikz}
\usepackage{multirow}
\usepackage{algorithm}
\usepackage{siunitx} 
\usepackage{algcompatible}
\usepackage{algpseudocode}
\usepackage{orcidlink}

\usepackage{hyperref}

\begin{document}
\title{WDM: 3D Wavelet Diffusion Models for High-Resolution Medical Image Synthesis}
\titlerunning{3D Wavelet Diffusion Models for High-Resolution Medical Image Synthesis}
%
\author{Paul Friedrich\orcidlink{0000-0003-3653-5624} \and 
        Julia Wolleb\orcidlink{0000-0003-4087-5920} \and 
        Florentin Bieder\orcidlink{0000-0001-9558-0623} \and 
        Alicia Durrer\orcidlink{0009-0007-8970-909X} \and 
        Philippe C. Cattin\orcidlink{0000-0001-8785-2713}}
%
\authorrunning{P. Friedrich et al.}

\institute{Department of Biomedical Engineering, University of Basel, Allschwil, Switzerland\\
\email{paul.friedrich@unibas.ch}}
\maketitle              
\begin{abstract}
Due to the three-dimensional nature of CT- or MR-scans, generative modeling of medical images is a particularly challenging task. Existing approaches mostly apply patch-wise, slice-wise, or cascaded generation techniques to fit the high-dimensional data into the limited GPU memory. However, these approaches may introduce artifacts and potentially restrict the model's applicability for certain downstream tasks. This work presents WDM, a wavelet-based medical image synthesis framework that applies a diffusion model on wavelet decomposed images. The presented approach is a simple yet effective way of scaling 3D diffusion models to high resolutions and can be trained on a single \SI{40}{\giga\byte} GPU. Experimental results on BraTS and LIDC-IDRI unconditional image generation at a resolution of $128 \times 128 \times 128$ demonstrate state-of-the-art image fidelity (FID) and sample diversity (MS-SSIM) scores compared to recent GANs, Diffusion Models, and Latent Diffusion Models. Our proposed method is the only one capable of generating high-quality images at a resolution of $256 \times 256 \times 256$, outperforming all comparing methods. The project page is available at \url{https://pfriedri.github.io/wdm-3d-io}.
\end{abstract}
\section{Introduction}
\begin{figure}
    \centering
    \includegraphics[width=\textwidth]{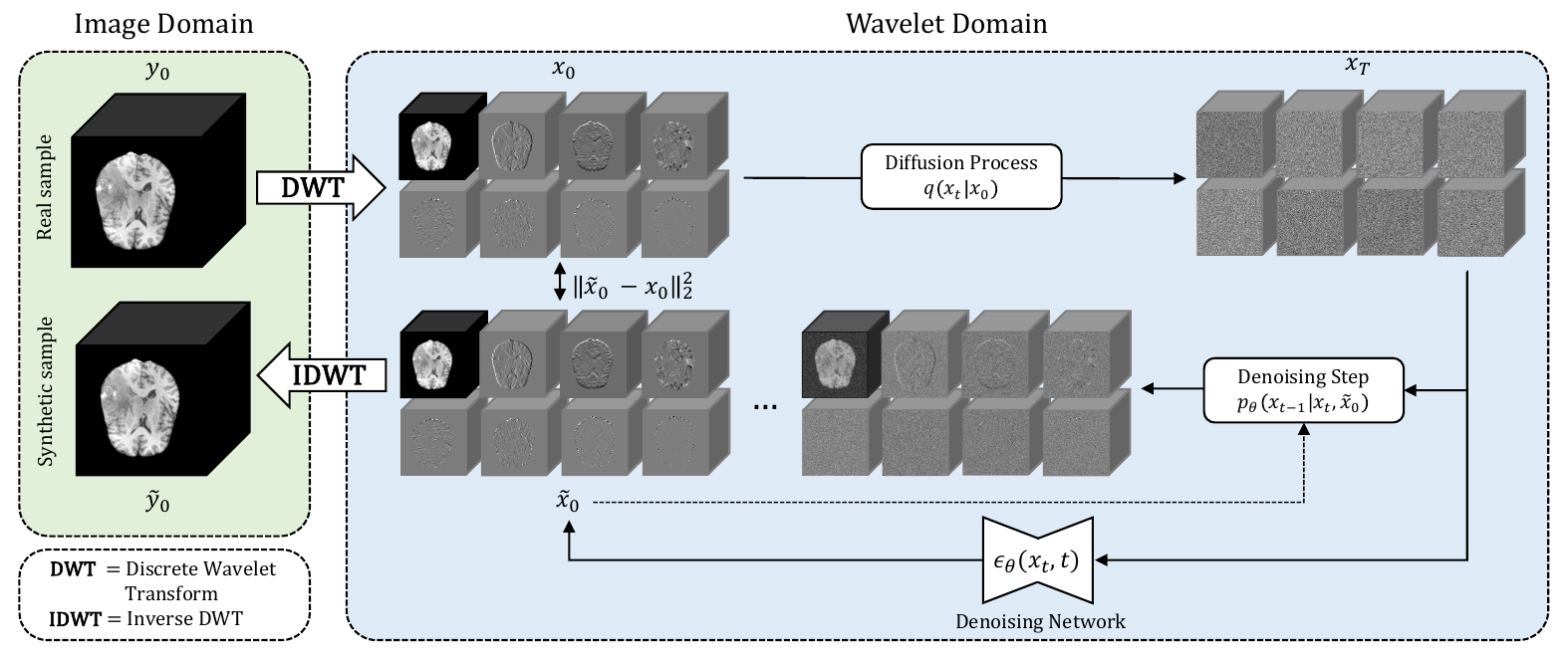}
    \caption{Schematic overview of the proposed wavelet-based image synthesis framework. A diffusion model is trained on the wavelet coefficients $x_0$ of the real input data $y_0$. During sampling, starting from random wavelet coefficients $x_T$, $T$ denoising steps are performed iteratively to predict denoised wavelet coefficients $\tilde{x}_0$. The final output images $\tilde{y}_0$ are produced by applying Inverse Discrete Wavelet Transform (IDWT) to the generated wavelet coefficients $\tilde{x}_0$.}
    \label{fig:overview}
\end{figure}

Costly and labor-intensive image acquisition, as well as privacy concerns, are reasons for one of the main challenges in medical image analysis with deep learning - the lack of large-scale datasets. Recent research \cite{ktena2023generative,sagers2023augmenting} suggests that generative models can help to close this gap by augmenting small datasets with synthetic data. In addition, generative models often form the basis for downstream tasks like semantic segmentation, anomaly detection, image reconstruction, or image-to-image translation. They are, therefore, of great value to the entire community. Due to the three-dimensional nature of medical images, like CT- or MR-scans, generative modeling of high-resolution medical images is particularly challenging and often limited by the available GPU memory. While recent research on this topic focused on scaling to high resolutions via cascaded image generation or by applying generative models on learned latent representations of the data, we present a different approach that applies diffusion models on wavelet-decomposed images. An overview of our method is shown in Fig. \ref{fig:overview}.
The proposed wavelet diffusion model, further called WDM, effectively reduces the required GPU memory and scales to high resolutions. Our main contributions are:
\begin{itemize}
    \item We propose WDM, a memory-efficient 3D Wavelet Diffusion Model for medical image synthesis.
    \item The proposed method can generate high-quality images with resolutions up to $256 \times 256 \times 256$ and can be trained on a single \SI{40}{\giga\byte} GPU.
    \item The proposed method demonstrates state-of-the-art image fidelity (FID) and sample diversity (MS-SSIM) scores at a resolution of $128 \times 128 \times 128$, while outperforming all comparing methods at a resolution of $256 \times 256 \times 256$.
\end{itemize}
\subsubsection{Related Work}
Previous research on 3D medical image synthesis mainly focused on Variational Autoencoders (VAEs) \cite{volokitin2020modelling} and Generative Adversarial Networks (GANs) \cite{ferreira2024gangen,hong20213d,kwon2019generation,yu20183d}. While high GPU memory consumption limits most of these approaches to a maximum resolution of $128 \times 128 \times 128$, Sun et al. \cite{sun2022hierarchical} presented a hierarchical amortized GAN capable of progressively generating images with a resolution up to $256 \times 256 \times 256$. GAN-based approaches, however, suffer from inherent issues such as training instability and mode collapse, restricting their applicability in certain scenarios. Diffusion probabilistic models \cite{ho2020denoising,sohl2015deep} have recently shown great performance in image generation \cite{dhariwal2021diffusion} and have widely been adopted for medical imaging tasks like semantic segmentation \cite{wolleb2022diffusion,wu2023medsegdiff}, anomaly detection \cite{wolleb2022anomaly,wyatt2022anno}, contrast harmonization \cite{durrer2023diffusion}, and implant generation \cite{friedrich2023point}. The application of diffusion models for 3D medical image generation has been explored in \cite{dorjsembe2022threedimensional,peng2023generating} and has already shown promising results. The available GPU memory, however, still remains the main limiting factor. To address this issue, latent diffusion models (LDMs) \cite{khader2023denoising,pinaya2022brain,rombach2022high,zhu2023make} that operate on a learned low-dimensional latent representation of the data have been proposed. They reduce the memory consumption and computational cost of the diffusion model but require a pretrained autoencoder. Training such an 3D autoencoder with a good latent representation is itself challenging for high-resolution volumes, and remains constrained by the available GPU memory. Instead of operating on a learned compressed latent representation of the input data, \cite{gal2021swagan,guth2022wavelet,phung2023wavelet} suggest using generative models on the wavelet coefficients of the input images. In this work, we eliminate the need for a first-stage autoencoder by adapting these ideas and combining them with recent advances in memory-efficient architectures to build a framework for high-resolution 3D medical image synthesis.
\section{Background}
\subsubsection{Diffusion Models} 
Denoising Diffusion Models \cite{ho2020denoising,nichol2021improved} are latent variable models that sample from a distribution by reversing a noising process. Given a sample $x_0$ from the real data distribution, this noising process progressively perturbs the sample with Gaussian noise for a defined number of timesteps $T$, following a sequence of normal distributions
\begin{equation}
    q(x_t|x_{t-1}) := \mathcal{N}(\sqrt{1-\beta_t}x_{t-1}, \beta_t\boldsymbol{I}),
\end{equation}
where $t \in \{1, ..., T\}$. This noising process has a predefined variance schedule $\beta_{1:T}$ and transforms the sample $x_0$ into a standard normal distribution $x_T \sim \mathcal{N}(0, \boldsymbol{I})$ for sufficiently large $T$. The reverse process is modeled as a Markov chain
\begin{equation}
    p_{\theta}(x_{0:T}) := p(x_T)\prod_{t=1}^Tp_{\theta}(x_{t-1}|x_t, \tilde{x}_0),
\end{equation}
with each transition being a Gaussian with mean $\mu_{t}(x_t, \tilde{x}_0)$ parameterized by a time-conditioned neural network $\epsilon_\theta$.
By defining $\alpha_t:=1-\beta_t$ and $\bar{\alpha}_t := \prod_{s=1}^t\alpha_s$, we can sample a noise perturbed latent $x_t$ at a timestep $t$
\begin{equation}
    q(x_t|x_0) := \mathcal{N}(\sqrt{\bar{\alpha}_t}x_0, (1-\bar{\alpha}_t)\boldsymbol{I}),
\end{equation}
and train $\epsilon_\theta$ to predict the denoised image $\tilde{x}_0 = \epsilon_\theta(x_t, t)$. We do this by applying an Mean Squared Error (MSE) loss between predicted $\tilde{x}_0$ and ground truth denoised image $x_0$
\begin{equation}
    \mathcal{L}_{MSE} = \|\tilde{x}_0 - x_0\|^{2}_{2}.
\end{equation}
For generating new samples, we draw $x_T \sim \mathcal{N}(0,\boldsymbol{I})$ and iteratively sample $x_{t-1}$ from the posterior distribution  
\begin{equation}
    p_\theta(x_{t-1}|x_t, \tilde{x}_0) = \mathcal{N}(\mu_t(x_t, \tilde{x}_0), \tilde{\beta}_t\boldsymbol{I})
\end{equation}
by plugging in the following equations for $t=T, ..., 1$:
\begin{equation}
   \mu_t(x_t, \tilde{x}_0) := \frac{\sqrt{\bar{\alpha}_t-1}\beta_t}{1-\bar{\alpha}_t}\tilde{x}_0 + \frac{\sqrt{\alpha_t}(1-\bar{\alpha}_{t-1})}{1-\bar{\alpha}_t}x_t,
\end{equation}
\begin{equation}
    \tilde{\beta}_t := \frac{1-\bar{\alpha}_{t-1}}{1-\bar{\alpha}_t}\beta_t.
\end{equation}
\subsubsection{Wavelet Transform}
The Discrete Wavelet Transform (DWT) is a widely used time-frequency analysis tool. It consists of a combination of low- and high-pass filters with stride 2, denoted as $l=\frac{1}{\sqrt{2}} \begin{bmatrix}1 & 1\end{bmatrix}$ and $h=\frac{1}{\sqrt{2}} \begin{bmatrix}-1 & 1\end{bmatrix}$, that are applied along all three spatial dimensions. They decompose a 3D volume ${y \in \mathbb{R}^{D \times H \times W}}$ into 8 wavelet coefficients $(x_{lll},x_{llh},x_{lhl},x_{lhh},x_{hll},x_{hlh},x_{hhl},x_{hhh}) = \text{DWT}(y)$ with half the spatial resolution $x_{\{lll, ..., hhh\}} \in \mathbb{R}^{\frac{D}{2} \times \frac{H}{2} \times \frac{W}{2}}$. By applying the Inverse Discrete Wavelet Transform (IDWT), we can obtain the original image $y$ from the wavelet coefficients $y = \text{IDWT}(x_{lll}, ..., x_{hhh})$. In this work, we apply wavelet transform to reduce the spatial dimension of the input volumes by aggregating the decomposed features in the channel dimension, offering a training-free alternative to commonly used compression with autoencoders. We apply Haar wavelets using an implementation presented in \cite{li2020wavelet}.
\section{Method}
\subsubsection{Wavelet-Based Image Synthesis}
We present a 3D medical image synthesis framework that produces high-resolution images by generating synthetic wavelet coefficients followed by IDWT. Given an input image $y \in \mathbb{R}^{D \times H \times W}$, we first apply DWT to decompose this image into its 8 wavelet coefficients. We concatenate them to form a single target matrix $x \in \mathbb{R}^{8 \times \frac{D}{2} \times \frac{H}{2} \times \frac{W}{2}}$ to be predicted by a diffusion model. When processing this matrix $x$, we first map it onto the network's base channels $\mathcal{C}$ (number of channels in the input layer) via a first convolution, leaving the network width unchanged compared to standard architectures. As our network then operates on the wavelet domain only, we profit from an $8 \times$ reduction in spatial dimension, allowing for shallower network architectures, less computations and a significantly reduced memory footprint.
\noindent\begin{minipage}{0.48\textwidth}
    \vfill
    \begin{algorithm}[H]
        \caption{Wavelet-based training}\label{alg:cap1}
        \begin{algorithmic}
            \Repeat\vphantom{$\|_{a}^{b}$}
            \State\vphantom{$\|_{a}^{b}$}$x_0 = \text{DWT}(y_0)$
            \State\vphantom{$\|_{a}^{b}$}$t \sim \text{Uniform}(\{1, ..., T\})$
            \State\vphantom{$\|_{a}^{b}$}$x_t \sim q(x_t|x_0)$
            \State\vphantom{$\|_{a}^{b}$}$\text{Perform gradient descent on}$
            \Statex\hspace{2em}\vphantom{$\|_{a}^{b}$}$\nabla_\theta \|\tilde{x}_0 - x_0\|^{2}_{2}, \text{where } \tilde{x}_0=\epsilon_\theta(x_t, t)$
            \Until{converged}\vphantom{$\|_{a}^{b}$}
        \end{algorithmic}
    \end{algorithm}
\vfill
\end{minipage}
\hfill
\begin{minipage}{0.48\textwidth}
    \vfill
    \begin{algorithm}[H]
        \caption{Wavelet-based sampling}\label{alg:cap2}
        \begin{algorithmic}
            \State\vphantom{$\|_{a}^{b}$}$x_T \sim \mathcal{N}(0,\boldsymbol{I})$
            \For{$t =T, ..., 1$}\vphantom{$\|_{a}^{b}$}
            \State\vphantom{$\|_{a}^{b}$}$\tilde{x}_0 = \epsilon_\theta(x_t, t)$
            \State\vphantom{$\|_{a}^{b}$}$x_{t-1} \sim p_\theta(x_{t-1}|x_t, \tilde{x}_0)$
            \EndFor\vphantom{$\|_{a}^{b}$}
            \State\vphantom{$\|_{a}^{b}$}$\tilde{y}_0 = \text{IDWT}(x_0)$
            \State\vphantom{$\|_{a}^{b}$}\textbf{return} $\tilde{y}_0$
        \end{algorithmic}
    \end{algorithm}
\vfill
\end{minipage}
During training, shown in Algorithm \ref{alg:cap1}, we draw a corrupted sample $x_t$ from $q(x_t|x_0)$ and train a diffusion model to generate an approximation $\tilde{x}_0$ of the original wavelet coefficients $x_0$. We also experimented with predicting the noise to be removed from corrupted wavelet coefficients, which led to the generation of images showing checkerboard artifacts. We train our network using an MSE loss between the predicted $\tilde{x}_0$ and the ground truth wavelet coefficients $x_0$. We further explored applying the loss function in the image domain, but found no improvements over a wavelet-level loss. During inference, shown in Algorithm \ref{alg:cap2}, we draw $x_T \sim \mathcal{N}(0,\boldsymbol{I})$ and pass it through the reverse diffusion process for $t=T, ..., 1$. The output image $\tilde{y}_0$ can finally be reconstructed by applying IDWT.
\section{Experiments}
\subsection{Experimental Settings}
\subsubsection{Datasets}
We demonstrate our model's performance on two publicly available datasets with different modalities: BraTS 2023 Adult Glioma \cite{baid2021rsna,bakas2017advancing,karargyris2023federated,menze2014multimodal} and LIDC-IDRI \cite{armato2011lung}. For training on the BraTS dataset, we use the T1-weighted brain MR-images, clip the upper and lower 0.1 percentile intensity values, zero pad the volumes to a size of $256 \times 256 \times 256$ and normalize them to a range of $[-1, 1]$. For training on the LIDC-IDRI lung CT dataset, we initially clip all values below $-1000$ and the upper 0.1 percentile intensity values. We then resample the images to an isotropic voxel size of 1 \si{\milli\metre}, center crop them to a size of $256 \times 256 \times 256$, and normalize to a range of $[-1, 1]$. When training on  images with a resolution of $128 \times 128 \times 128$, we perform downsampling by applying average pooling.
\subsubsection{Evaluation Metrics}
We use the Fréchet Inception Distance (FID) and the Multi-Scale Structural Similarity Index Measure (MS-SSIM) to assess the generated images' fidelity and diversity. Following \cite{pinaya2022brain} and \cite{sun2022hierarchical}, we compute FID scores over 1k samples using a pretrained Med3D network \cite{chen2019med3d}. A low FID score suggests that the generated images' distribution is close to the distribution of real samples, meaning that the generated images appear realistic. Following \cite{pinaya2022brain}, we evaluate the generated images' diversity by measuring the mean MS-SSIM over 1k generated images, where a low MS-SSIM score indicates a high sample diversity. We also measure the required GPU memory during inference.
\subsubsection{Implementation Details}
We build our model $\epsilon_\theta$ on a memory efficient 3D diffusion model architecture with additive skip-connections \cite{bieder2023memory}. We define a diffusion process with $T=1000$ timesteps, and a linear variance schedule between $\beta_1 = 1 \times 10^{-4}$ and $\beta_T=0.02$. We set our model's number of base channels to $\mathcal{C}=64$ for all experiments, use the Adam optimizer with a learning rate of $1 \times 10^{-5}$, and a batch size of $10/1$ for images with a resolution of $128^{3}/256^{3}$ respectively. We train our models for \SI{1.2}{M} iterations on the low and for \SI{2}{M} iterations on the high-resolution images. All experiments were carried out on a single NVIDIA A100 (\SI{40}{\giga\byte}) GPU. A detailed list of all hyperparameters can be found in the Supplementary Material. The code, as well as implementation details for comparing methods, are available at \url{https://github.com/pfriedri/wdm-3d}.
\subsubsection{Wavelet-Informed Network Architecture (WavU-Net)} Recent research on 2D natural image generation \cite{phung2023wavelet} showed that incorporating wavelet information into the U-Net feature space can be beneficial to the overall image quality. To evaluate whether this also applies to 3D medical images, we additionally evaluate an architecture with wavelet up- and downsampling operations and wavelet residual connections. We refer to this architecture of $\epsilon_\theta$ as WavU-Net. An overview of the architecture is shown in the Supplementary Material.
\subsection{Experimental Results}
\subsubsection{Comparing Methods} We evaluate our method WDM on an unconditional image generation task and compare it with: (1) HA-GAN \cite{sun2022hierarchical}, a hierarchical 3D GAN approach that progressively generates synthetic data by sampling low-resolution images and performing a learned upsampling operation; (2) 3D LDM \cite{khader2023denoising}, a 3D diffusion model trained on the learned latent space of a 3D VQ-GAN; (3) 2.5D LDM \cite{zhu2023make}, a 2D slice-wise diffusion model trained on the learned latent space of a 2D VQ-VAE, augmented with volumetric layers to improve slice-consistency; (4) 3D DDPM, a memory-efficient 3D diffusion model implementation from \cite{bieder2023memory}. By using the same hyperparameters as our method, 3D DDPM serves as a baseline diffusion model operating on the image domain only.
\subsubsection{Results} 
\begin{table}
    \centering
    \caption{Results on BraTS unconditional brain MR image generation. FID scores are multiplied by $10^3$. "---" indicates that the method can not be trained on a single \SI{40}{\giga\byte} GPU. A * indicates our adopted version for unconditional image generation.}
    \begin{tabular}{lcccccc}
        \multirow{2}{*}{\textbf{Method}}  & \multicolumn{3}{c}{$128 \times 128 \times 128$} & \multicolumn{3}{c}{$256 \times 256 \times 256$}\\
        & FID $\downarrow$ & MS-SSIM $\downarrow$ & Mem (GB) $\downarrow$ & FID $\downarrow$ & MS-SSIM $\downarrow$ & Mem (GB) $\downarrow$ \\ \hline
        \bf{WDM}                            & 0.154 & 0.888 & 2.55 & 0.379 & 0.890 & 7.27 \\
        WDM (WavU-Net)                      & 0.259 & 0.879 & 2.65 & 0.656 & 0.860 & 7.59 \\\hline
        HA-GAN \cite{sun2022hierarchical}   & 0.785 & 0.905 & 2.58 & 1.368 & 0.999 & 6.01 \\
        3D LDM \cite{khader2023denoising}   & 1.394 & 0.926 & 9.82 & ---   & ---   & ---  \\
        2.5D LDM * \cite{zhu2023make}       & 81.06 & 0.579 & 6.81 & ---   & ---   & ---  \\
        3D DDPM                             & 1.402 & 0.876 & 6.51 & ---   & ---   & ---  \\
    \end{tabular}
    \label{tab:brats}
\end{table}
\begin{table}
    \centering
    \caption{Results on LIDC-IDRI unconditional lung CT image generation. "---" indicates that the method can not be trained on a single \SI{40}{\giga\byte} GPU. A * indicates our adopted version for unconditional image generation.}
    \begin {tabular}{lcccccc}
        \multirow{2}{*}{\textbf{Method}}  & \multicolumn{3}{c}{$128 \times 128 \times 128$} & \multicolumn{3}{c}{$256 \times 256 \times 256$}\\
        & FID $\downarrow$ & MS-SSIM $\downarrow$ & Mem (GB) $\downarrow$ & FID $\downarrow$ & MS-SSIM $\downarrow$ & Mem (GB) $\downarrow$ \\\hline
        \bf{WDM}                            & 4.989 & 0.306 & 2.55 & 4.176 & 0.270 & 7.27 \\
        WDM (WavU-Net)                      & 5.019 & 0.310 & 2.65 & 4.121 & 0.271 & 7.59 \\\hline
        HA-GAN \cite{sun2022hierarchical}   & 5.002 & 0.371 & 2.58 & 4.224 & 0.357 & 6.01 \\
        3D LDM \cite{khader2023denoising}   & 6.083 & 0.377 & 9.82 & ---   & ---   & ---  \\
        2.5D LDM * \cite{zhu2023make}       & 5.060 & 0.280 & 6.81 & ---   & ---   & ---  \\
        3D DDPM                             & 4.850 & 0.266 & 6.51 & ---   & ---   & ---  \\
    \end{tabular}
    \label{tab:lidc-idri}
\end{table}
The results presented in Table \ref{tab:brats} and \ref{tab:lidc-idri} show our proposed method's performance on an unconditional 3D brain MR and chest CT image generation task. Our approach not only outperforms current state-of-the-art methods in FID and MS-SSIM metrics, it also has the lowest inference GPU memory footprint at a resolution of $128 \times 128 \times 128$ and is the only diffusion-based method that can be trained at a resolution of $256 \times 256 \times 256$. Operating in the wavelet domain and profiting from the reduced spatial dimension results in relatively short inference times of \SI{35}{\second}/\SI{240}{\second} at the respective resolutions. Compared to the results presented in \cite{phung2023wavelet}, we do not find that incorporating wavelet information into the network's feature space (WavU-Net) increases the model's performance on 3D medical volumes. Qualitative results of our proposed method (WDM) are shown in Fig. \ref{fig:brats} and \ref{fig:lidc}. A qualitative comparison of samples produced by all evaluated methods can be found in the Supplementary Material.
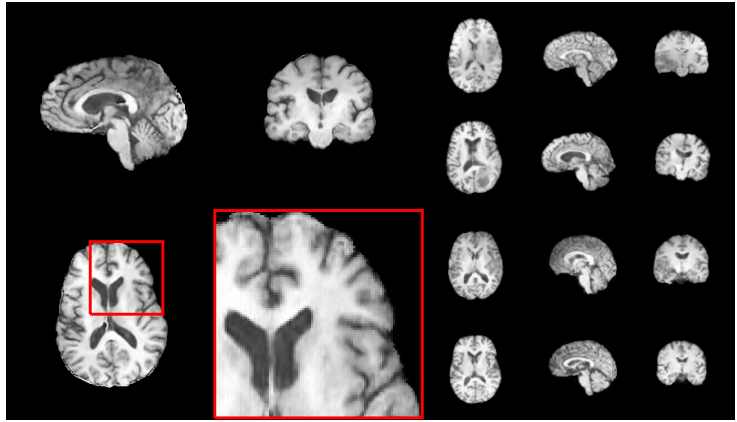
\begin{figure}
    \centering
    \resizebox{.84\textwidth}{!}{
        \begin{tikzpicture} 
            \fill (0.1, 0.1) rectangle (7.125, 4.125);
        	\node[] at (0, 0)[anchor=south west]  {\includegraphics[height=2cm]{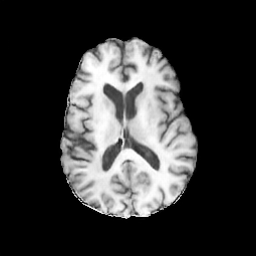}};
            \node[] at (0, 2)[anchor=south west]  {\includegraphics[height=2cm]{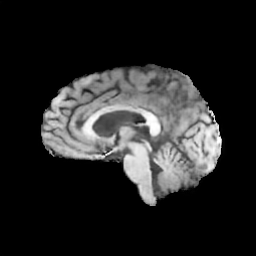}};
            \node[] at (2, 2)[anchor=south west]  {\includegraphics[height=2cm]{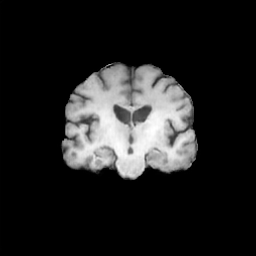}};
            \node[] at (2, 0)[anchor=south west]  {\includegraphics[height=2cm]{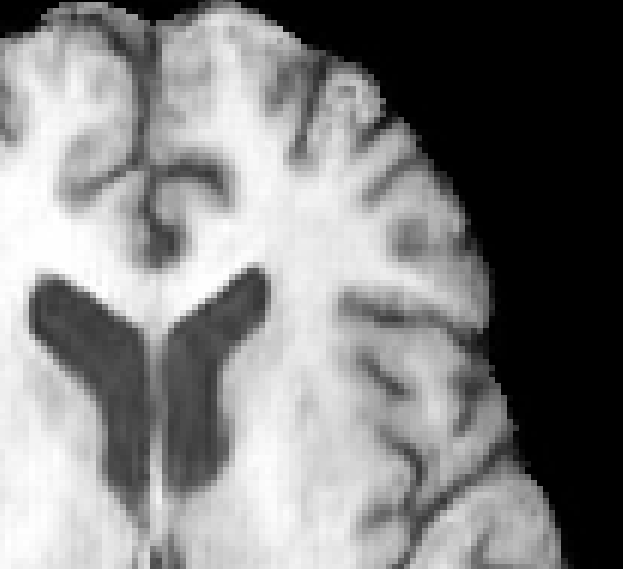}};
            \node[] at (4, 3)[anchor=south west]  {\includegraphics[height=1cm]{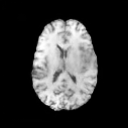}};
            \node[] at (5, 3)[anchor=south west]  {\includegraphics[height=1cm]{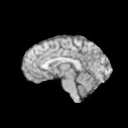}};
            \node[] at (6, 3)[anchor=south west]  {\includegraphics[height=1cm]{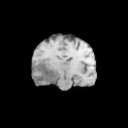}};
            \node[] at (4, 2)[anchor=south west]  {\includegraphics[height=1cm]{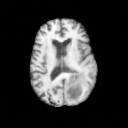}};
            \node[] at (5, 2)[anchor=south west]  {\includegraphics[height=1cm]{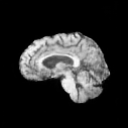}};
            \node[] at (6, 2)[anchor=south west]  {\includegraphics[height=1cm]{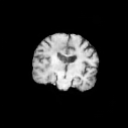}};
            \node[] at (4, 1)[anchor=south west]  {\includegraphics[height=1cm]{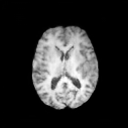}};
            \node[] at (5, 1)[anchor=south west]  {\includegraphics[height=1cm]{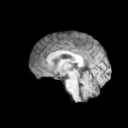}};
            \node[] at (6, 1)[anchor=south west]  {\includegraphics[height=1cm]{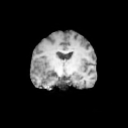}};
            \node[] at (4, 0)[anchor=south west]  {\includegraphics[height=1cm]{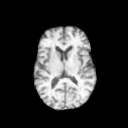}};
            \node[] at (5, 0)[anchor=south west]  {\includegraphics[height=1cm]{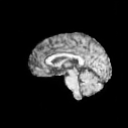}};
            \node[] at (6, 0)[anchor=south west]  {\includegraphics[height=1cm]{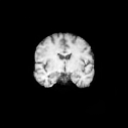}};
            \draw[thick,red] (2.115,0.115) -- (4.115,0.115) -- (4.115,2.115) -- (2.115,2.115) -- (2.115,0.115);
            \draw[thick,red] (0.915,1.115) -- (0.915,1.815) -- (1.615,1.815) -- (1.615,1.115) -- (0.915,1.115);
    	\end{tikzpicture}
    }
    \caption{Qualitative results of our method (WDM) on an unconditional image generation task on BraTS $256 \times 256 \times 256$ \textit{(left)} and $128 \times 128 \times 128$ \textit{(right)}.}
    \label{fig:brats}
\end{figure}
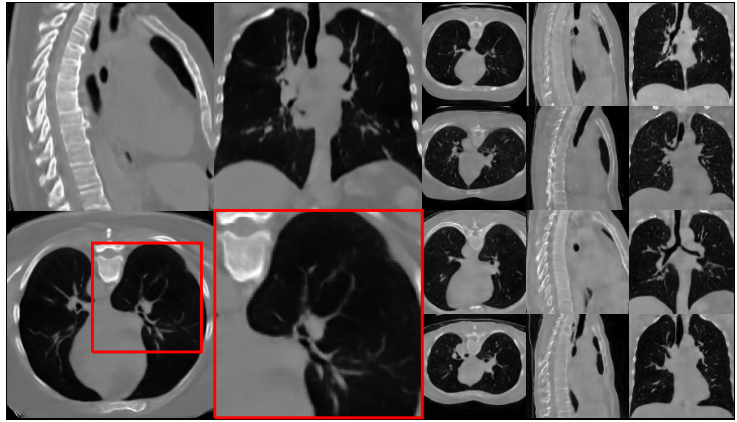
\begin{figure}
    \centering
    \resizebox{.84\textwidth}{!}{
        \begin{tikzpicture} 
            \fill (0.1, 0.1) rectangle (7.125, 4.125);
        	\node[] at (0, 0)[anchor=south west]  {\includegraphics[height=2cm]{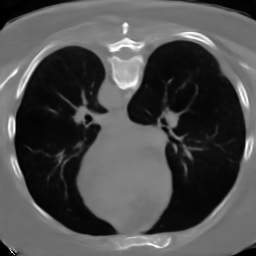}};
            \node[] at (0, 2)[anchor=south west]  {\includegraphics[height=2cm]{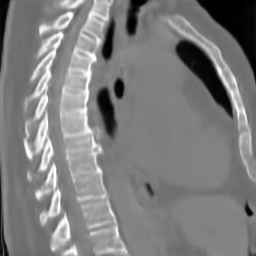}};
            \node[] at (2, 2)[anchor=south west]  {\includegraphics[height=2cm]{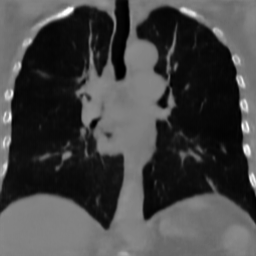}};
            \node[] at (2, 0)[anchor=south west]  {\includegraphics[height=2cm]{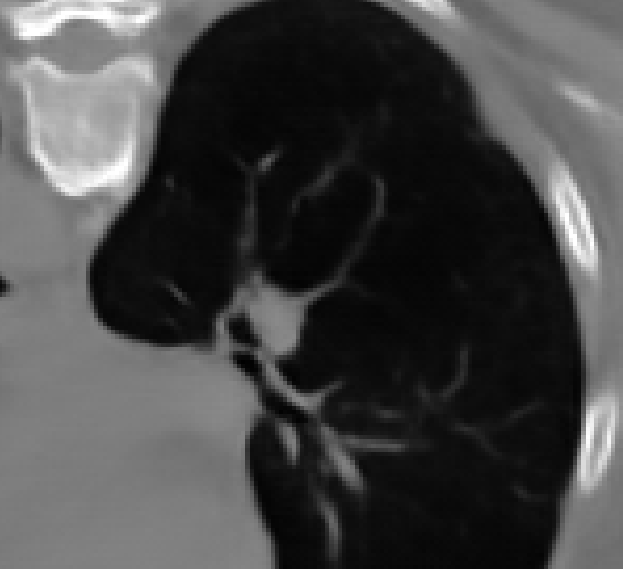}};
            \node[] at (4, 3)[anchor=south west]  {\includegraphics[height=1cm]{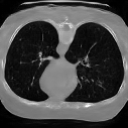}};
            \node[] at (5, 3)[anchor=south west]  {\includegraphics[height=1cm]{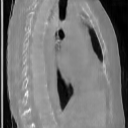}};
            \node[] at (6, 3)[anchor=south west]  {\includegraphics[height=1cm]{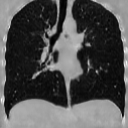}};
            \node[] at (4, 2)[anchor=south west]  {\includegraphics[height=1cm]{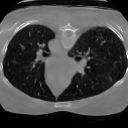}};
            \node[] at (5, 2)[anchor=south west]  {\includegraphics[height=1cm]{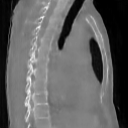}};
            \node[] at (6, 2)[anchor=south west]  {\includegraphics[height=1cm]{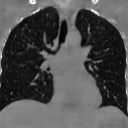}};
            \node[] at (4, 1)[anchor=south west]  {\includegraphics[height=1cm]{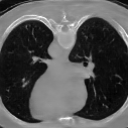}};
            \node[] at (5, 1)[anchor=south west]  {\includegraphics[height=1cm]{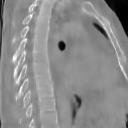}};
            \node[] at (6, 1)[anchor=south west]  {\includegraphics[height=1cm]{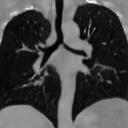}};
            \node[] at (4, 0)[anchor=south west]  {\includegraphics[height=1cm]{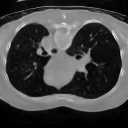}};
            \node[] at (5, 0)[anchor=south west]  {\includegraphics[height=1cm]{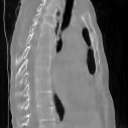}};
            \node[] at (6, 0)[anchor=south west]  {\includegraphics[height=1cm]{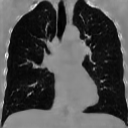}};
            \draw[thick,red] (2.115,0.115) -- (4.115,0.115) -- (4.115,2.115) -- (2.115,2.115) -- (2.115,0.115);
            \draw[thick,red] (0.9359,0.75) -- (0.9359,1.8) -- (1.9859,1.8) -- (1.9859,0.75) -- (0.9359,0.75);
    	\end{tikzpicture}
    }
    \caption{Qualitative results of our method (WDM) on an unconditional image generation task on LIDC-IDRI $256 \times 256 \times 256$ \textit{(left)} and $128 \times 128 \times 128$ \textit{(right)}.}
    \label{fig:lidc}
\end{figure}
\subsubsection{Discussion} (1) Our proposed method outperforms HA-GAN \cite{sun2022hierarchical} regarding both FID and MS-SSIM metrics. While HA-GAN offers a slightly reduced memory footprint at higher resolutions, we find that it tends to suffer from severe mode collapse, indicated by a high MS-SSIM score of $0.999$ on BraTS. As HA-GAN does not require iterative sampling, it is the method with the shortest inference time of \SI{0.4}{\second}/\SI{3}{\second} at the respective resolutions.
(2) Our proposed method outperforms 3D LDM \cite{khader2023denoising} on all assessed metrics. Since 3D LDM generates images on a learned latent representation of the original data, its performance heavily depends on the quality of the 3D autoencoder, the component that also limits the method to the lower resolution. It would be possible to train a smaller autoencoder and trade image quality for resolution, which does not seem optimal. 3D LDM has a relatively short inference time of \SI{8}{\second} for a diffusion process with $T=300$.
(3) Our proposed method outperforms 2.5D LDM \cite{zhu2023make} in terms of FID. We find that this pseudo-3D method is not able to generate meaningful results on BraTS, which might be caused by a lot of background present in the images. Furthermore, it shows slice inconsistencies on LIDC-IDRI. This explains the low MS-SSIM score, caused by very diverse but unrealistic images. As originally proposed by the authors, the model seems to be suitable for image-to-image translation tasks conditioned on rich context information. The results for unconditional image generation remain insufficient. The method has an inference time of about \SI{52}{\second}.
(4) Compared to WDM, 3D DDPM offers similar FID and MS-SSIM scores while having a significantly increased GPU memory footprint. As it directly operates on the full resolution volumes, we were just able to train the method at a resolution of $128 \times 128 \times 128$. The inference time of about \SI{235}{\second} is the highest of all comparing methods. 

\noindent While we could try to further improve our proposed method's performance by using the gained GPU memory for applying larger networks, we demonstrate how wavelet transform can be used to effectively scale 3D diffusion models to high resolutions, save GPU memory and shorten the sampling time while maintaining the same standard network architecture as 3D DDPM. Compared to LDMs, our proposed wavelet-based approach offers a simple and dataset-agnostic tool for spatial dimensionality reduction without requiring additional training.
\section{Conclusion} This paper proposes WDM, an image synthesis framework for efficient high-resolution medical image generation, which applies a diffusion model to wavelet decomposed images. The presented approach outperforms state-of-the-art methods on an unconditional image generation task and effectively scales diffusion-based 3D medical image generation to high resolutions. Additionally, WDM has the potential to serve as an effective backbone, enabling the application of generation based medical downstream tasks to high-resolution volumes. For future work, we plan to extend our framework for conditional image generation, image inpainting, and image-to-image translation. We also plan to train our model as a Denoising Diffusion GAN \cite{xiao2022tackling} to further reduce the training and inference time.
\begin{credits}
\subsubsection{\ackname} This work was financially supported by the Werner Siemens Foundation through the MIRACLE II project.
\end{credits}
\bibliographystyle{splncs04.bst}
\bibliography{bibliography.bib}
\end{document}


\title{Supplementary Material}
\author{}
\institute{}
\maketitle
\section{Network Configuration}
\begin{table}
    \caption{Hyperparameters used to generate the results presented in the paper.}
    \centering
    \begin{tabular}{|l|c|c|c|c|}
    \hline
    Dataset & \multicolumn{2}{c|}{BraTS} & \multicolumn{2}{c|}{LIDC-IDRI}\\\hline
    Resolution                              & $128$                 & $256$              & $128$                & $256$ \\
    Number of base channels $\mathcal{C}$   & 64                    & 64                 & 64                   & 64 \\
    Number of ResBlocks per scale           & 2                     & 2                  & 2                    & 2 \\
    Channel multiplier per scale            & $(1,2,2,4,4)$         & $(1,2,2,4,4,4)$    & $(1,2,2,4,4)$        & $(1,2,2,4,4,4)$ \\
    Learning rate                           & $1 \times 10^{-5}$    & $1 \times 10^{-5}$ & $1 \times 10^{-5}$   & $1 \times 10^{-5}$ \\
    Batch size                              & 10                    & 1                  & 10                   & 1 \\
    Number of training iterations           & $1.2 \times 10^{6}$   & $2 \times 10^{6}$  & $1.2 \times 10^{6}$  & $2 \times 10^{6}$ \\
    Diffusion timesteps $T$                 & 1000                  & 1000               & 1000                 & 1000 \\
    Noise schedule                          & linear                & linear             & linear               & linear \\\hline
    \end{tabular}
    \label{tab:my_label}
\end{table}
\section{Detailed Architecture of Wavelet-Informed U-Net}
\begin{figure}
    \centering
    \includegraphics[width=\textwidth]{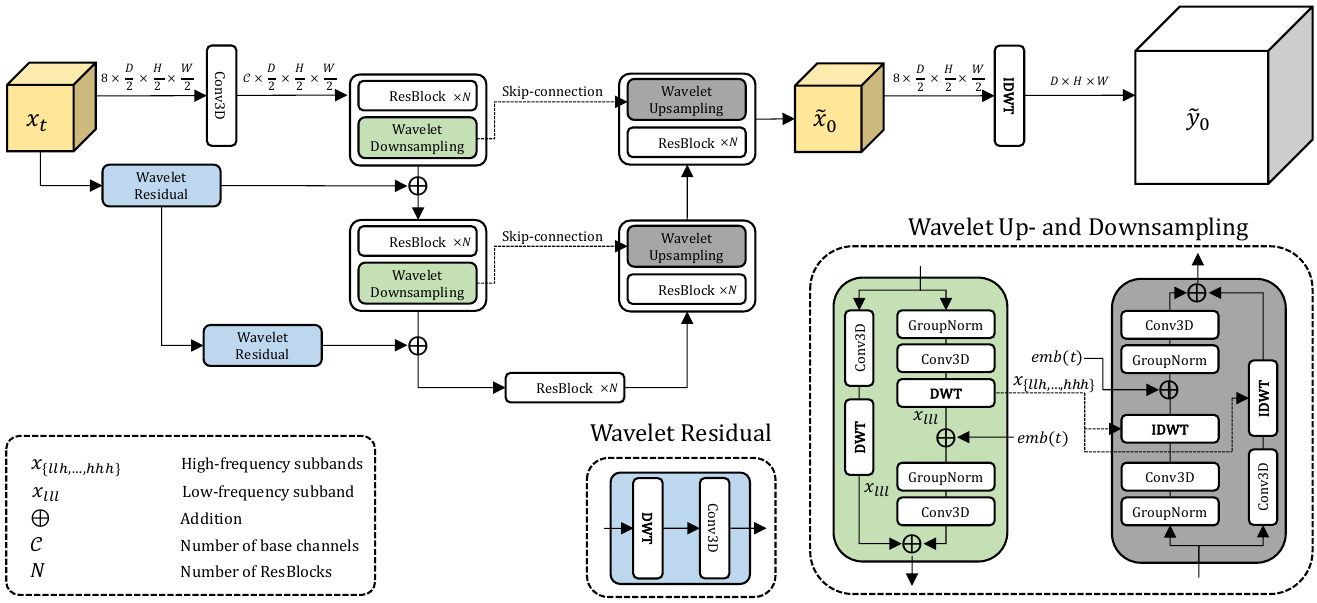}
    \caption{Overview of the Wavelet-Informed U-Net (WavU-Net). For simplicity, the timestep embedding $emb(t)$ of normal ResBlocks is missing. Our standard architecture looks similar, but does not use wavelet residuals, or wavelet up- and downsampling.}
    \label{fig:wavu-net}
\end{figure}
\section{Visual Comparison}
\begin{figure}[H]
    \centering
    \resizebox{.875\textwidth}{!}{
        \begin{tikzpicture} 
            \fill (0.1, 0.1) rectangle (6.125, 10.125);

            \node[] at (1.5, 10.25)       {\tiny BraTS};
            \node[] at (4.5, 10.25)       {\tiny LIDC-IDRI};

            \node[rotate=90] at (-0.125, 9)           {\tiny\textbf{WDM (Ours)}};
            \node[] at (0, 9)   [anchor=south west]  {\includegraphics[height=1cm]{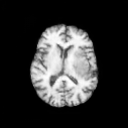}};
            \node[] at (1, 9)   [anchor=south west]  {\includegraphics[height=1cm]{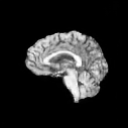}};
            \node[] at (2, 9)   [anchor=south west]  {\includegraphics[height=1cm]{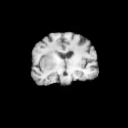}};
            \node[] at (0, 8)   [anchor=south west]  {\includegraphics[height=1cm]{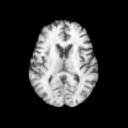}};
            \node[] at (1, 8)   [anchor=south west]  {\includegraphics[height=1cm]{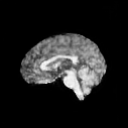}};
            \node[] at (2, 8)   [anchor=south west]  {\includegraphics[height=1cm]{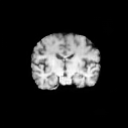}};
            \node[] at (3, 9)   [anchor=south west]  {\includegraphics[height=1cm]{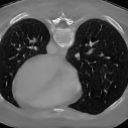}};
            \node[] at (4, 9)   [anchor=south west]  {\includegraphics[height=1cm]{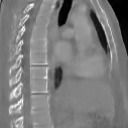}};
            \node[] at (5, 9)   [anchor=south west]  {\includegraphics[height=1cm]{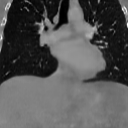}};
            \node[] at (3, 8)   [anchor=south west]  {\includegraphics[height=1cm]{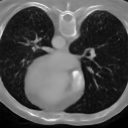}};
            \node[] at (4, 8)   [anchor=south west]  {\includegraphics[height=1cm]{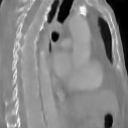}};
            \node[] at (5, 8)   [anchor=south west]  {\includegraphics[height=1cm]{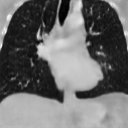}};

            \node[rotate=90] at (-0.125, 7)           {\tiny HA-GAN};
        	\node[] at (0, 7)   [anchor=south west]  {\includegraphics[height=1cm]{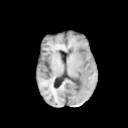}};
            \node[] at (1, 7)   [anchor=south west]  {\includegraphics[height=1cm]{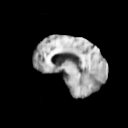}};
            \node[] at (2, 7)   [anchor=south west]  {\includegraphics[height=1cm]{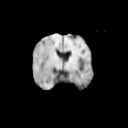}};
            \node[] at (0, 6)   [anchor=south west]  {\includegraphics[height=1cm]{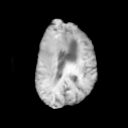}};
            \node[] at (1, 6)   [anchor=south west]  {\includegraphics[height=1cm]{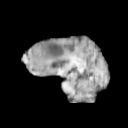}};
            \node[] at (2, 6)   [anchor=south west]  {\includegraphics[height=1cm]{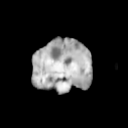}};
            \node[] at (3, 7)   [anchor=south west]  {\includegraphics[height=1cm]{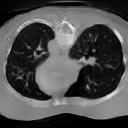}};
            \node[] at (4, 7)   [anchor=south west]  {\includegraphics[height=1cm]{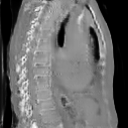}};
            \node[] at (5, 7)   [anchor=south west]  {\includegraphics[height=1cm]{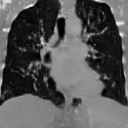}};
            \node[] at (3, 6)   [anchor=south west]  {\includegraphics[height=1cm]{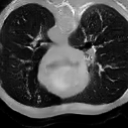}};
            \node[] at (4, 6)   [anchor=south west]  {\includegraphics[height=1cm]{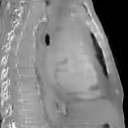}};
            \node[] at (5, 6)   [anchor=south west]  {\includegraphics[height=1cm]{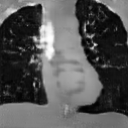}};

            \node[rotate=90] at (-0.125, 5)           {\tiny 3D LDM};
            \node[] at (0, 5)   [anchor=south west]  {\includegraphics[height=1cm]{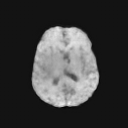}};
            \node[] at (1, 5)   [anchor=south west]  {\includegraphics[height=1cm]{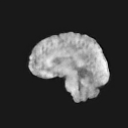}};
            \node[] at (2, 5)   [anchor=south west]  {\includegraphics[height=1cm]{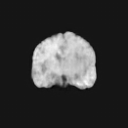}};
            \node[] at (0, 4)   [anchor=south west]  {\includegraphics[height=1cm]{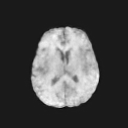}};
            \node[] at (1, 4)   [anchor=south west]  {\includegraphics[height=1cm]{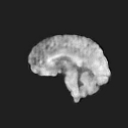}};
            \node[] at (2, 4)   [anchor=south west]  {\includegraphics[height=1cm]{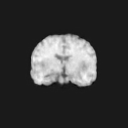}};
            \node[] at (3, 5)   [anchor=south west]  {\includegraphics[height=1cm]{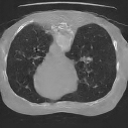}};
            \node[] at (4, 5)   [anchor=south west]  {\includegraphics[height=1cm]{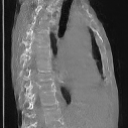}};
            \node[] at (5, 5)   [anchor=south west]  {\includegraphics[height=1cm]{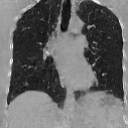}};
            \node[] at (3, 4)   [anchor=south west]  {\includegraphics[height=1cm]{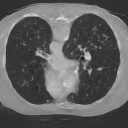}};
            \node[] at (4, 4)   [anchor=south west]  {\includegraphics[height=1cm]{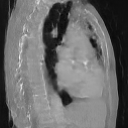}};
            \node[] at (5, 4)   [anchor=south west]  {\includegraphics[height=1cm]{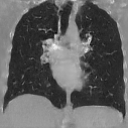}};

            \node[rotate=90] at (-0.125, 3)           {\tiny 2.5D LDM};
            \node[] at (0, 3)   [anchor=south west]  {\includegraphics[height=1cm]{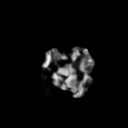}};
            \node[] at (1, 3)   [anchor=south west]  {\includegraphics[height=1cm]{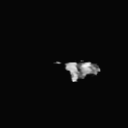}};
            \node[] at (2, 3)   [anchor=south west]  {\includegraphics[height=1cm]{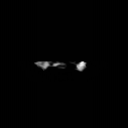}};
            \node[] at (0, 2)   [anchor=south west]  {\includegraphics[height=1cm]{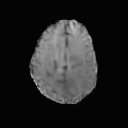}};
            \node[] at (1, 2)   [anchor=south west]  {\includegraphics[height=1cm]{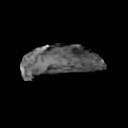}};
            \node[] at (2, 2)   [anchor=south west]  {\includegraphics[height=1cm]{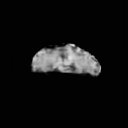}};
            \node[] at (3, 3)   [anchor=south west]  {\includegraphics[height=1cm]{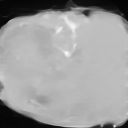}};
            \node[] at (4, 3)   [anchor=south west]  {\includegraphics[height=1cm]{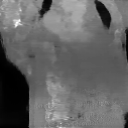}};
            \node[] at (5, 3)   [anchor=south west]  {\includegraphics[height=1cm]{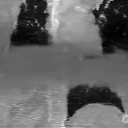}};
            \node[] at (3, 2)   [anchor=south west]  {\includegraphics[height=1cm]{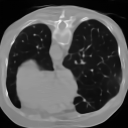}};
            \node[] at (4, 2)   [anchor=south west]  {\includegraphics[height=1cm]{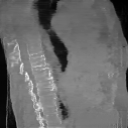}};
            \node[] at (5, 2)   [anchor=south west]  {\includegraphics[height=1cm]{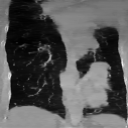}};

            \node[rotate=90] at (-0.125, 1)           {\tiny 3D DDPM};
            \node[] at (0, 1)   [anchor=south west]  {\includegraphics[height=1cm]{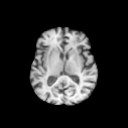}};
            \node[] at (1, 1)   [anchor=south west]  {\includegraphics[height=1cm]{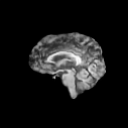}};
            \node[] at (2, 1)   [anchor=south west]  {\includegraphics[height=1cm]{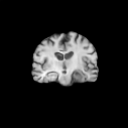}};
            \node[] at (0, 0)   [anchor=south west]  {\includegraphics[height=1cm]{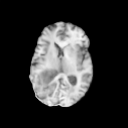}};
            \node[] at (1, 0)   [anchor=south west]  {\includegraphics[height=1cm]{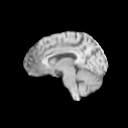}};
            \node[] at (2, 0)   [anchor=south west]  {\includegraphics[height=1cm]{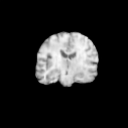}};
            \node[] at (3, 1)   [anchor=south west]  {\includegraphics[height=1cm]{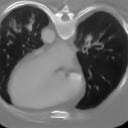}};
            \node[] at (4, 1)   [anchor=south west]  {\includegraphics[height=1cm]{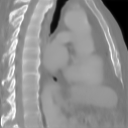}};
            \node[] at (5, 1)   [anchor=south west]  {\includegraphics[height=1cm]{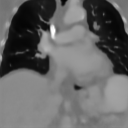}};
            \node[] at (3, 0)   [anchor=south west]  {\includegraphics[height=1cm]{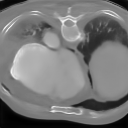}};
            \node[] at (4, 0)   [anchor=south west]  {\includegraphics[height=1cm]{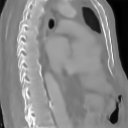}};
            \node[] at (5, 0)   [anchor=south west]  {\includegraphics[height=1cm]{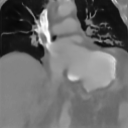}};
    	\end{tikzpicture}
    }
    \caption{A qualitative comparison of all evaluated methods on an unconditional brain MR and chest CT image generation task at a resolution of $128 \times 128 \times 128$.}
\end{figure}